%% file: BAM87_Ycp_resvised.tex
\documentclass[preprint,3p]{elsarticle}
\usepackage{graphics}
\usepackage{amssymb}
\usepackage{amsthm}
\usepackage[mathlines]{lineno}
\usepackage{graphicx}
\usepackage{units}
\usepackage{url}
\usepackage{amsmath}
\usepackage{amsfonts}
\usepackage{bm}
\usepackage{textcomp}
\usepackage{subfigure}
\usepackage{multicol}
\usepackage{verbatim}
\usepackage{booktabs}
\usepackage{multirow}
\usepackage{xspace}
\usepackage[colorlinks=true,linkcolor=blue]{hyperref}

\journal{Physics Letters B}

\newcommand{\ee}{\ensuremath{e^+e^-}\xspace}
\newcommand{\dbar}{\ensuremath{\overline{D}}\xspace}
\newcommand{\dzero}{\ensuremath{D^0}\xspace}
\newcommand{\dbarzero}{\ensuremath{\dbar{}^{0}}\xspace}
\newcommand{\gevcc}{\,\unit{GeV}/{c}^2}
\newcommand{\br}[1]{{\mathcal{B}}_{#1}}

\newcommand{\ycp}{y_{CP}}
\newcommand{\mbc}{M_\mathrm{BC}}
\newcommand{\umiss}{U_\mathrm{miss}}

\newcommand{\delE}{\Delta E}
\newcommand{\dEdx}{\ensuremath{\mathrm{d}E/\mathrm{d}x}\xspace}

\begin{document}
\begin{frontmatter}

\title{Measurement of $y_{CP}$ in $D^0-\overline{D}{}^0$ oscillation using quantum correlations in $e^+e^-\to D^0\overline{D}{}^0$ at $\sqrt{s}$ = 3.773\,GeV}

\input{authors_jan2015}

\vspace{0.4cm}


\begin{abstract}
We report a measurement of the parameter $\ycp$ in $\dzero-\dbarzero$
oscillations performed by taking advantage of quantum coherence
between pairs of $\dzero\dbarzero$ mesons produced in $e^+e^-$
annihilations near threshold. In this work, doubly-tagged
$\dzero\dbarzero$ events, where one $D$ decays to a $CP$ eigenstate
and the other $D$ decays in a semileptonic mode, are reconstructed
using a data sample of 2.92\,fb$^{-1}$ collected with the BESIII
detector at the center-of-mass energy of $\sqrt{s}$ = 3.773\,GeV.
We obtain $\ycp = (-2.0\pm1.3\pm0.7)\%$, where the first uncertainty
is statistical and the second is systematic.
This result is compatible with the current world average.
\end{abstract}


\end{frontmatter}

\section{Introduction}
\subsection{Charm oscillation}
It is well known that oscillations between meson and antimeson, also
called mixing, can occur when the flavor eigenstates differ from the
physical mass eigenstates.  These effects provide a mechanism whereby
interference in the transition amplitudes of mesons and antimesons may
occur.  They may also allow for manifestation of $CP$ violation
($CPV$) in the underlying dynamics~\cite{Bianco,xingzz}.  Oscillations
in the $K^{0}-\overline{K}{}^{0}$~\cite{KBBs1},
$B^{0}-\overline{B}{}^{0}$~\cite{KBBs2} and
$B_{s}^{0}-\overline{B}{}_{s}^{0}$~\cite{KBBs3} systems are
established; their oscillation rates are well-measured and consistent
with predictions from the standard model (SM)~\cite{PDG}.  After an
accumulation of strong evidence from a variety of
experiments~\cite{Staric:2007dt,Aubert:2007wf,Aaltonen:2007ac},
$\dzero-\dbarzero$ oscillations were recently firmly established by
LHCb~\cite{LHCb}.  The results were soon confirmed by CDF~\cite{CDF}
and Belle~\cite{Belle}.

The oscillations are conventionally characterized by two dimensionless
parameters $x=\Delta m/ \Gamma$ and $y=\Delta\Gamma/2\Gamma$, where
$\Delta m$ and $\Delta\Gamma$ are the mass and width differences
between the two mass eigenstates and $\Gamma$ is the average decay
width of those eigenstates.  The mass eigenstates can be written as
$|D_{1,2}\rangle = p|D^{0}\rangle \pm q|\overline{D}{}^{0}\rangle$,
where $p$ and $q$ are complex parameters and $\phi = \arg(q/p)$ is a
$CP$-violating phase. Using the phase convention
$CP|D^0\rangle=+|\dbarzero\rangle$, the $CP$ eigenstates of the $D$
meson can be written as
\begin{linenomath*}
\begin{equation}
|D_{CP+}\rangle\equiv\frac{|D^0\rangle+|\dbarzero\rangle}{\sqrt{2}},~~~|D_{CP-}\rangle\equiv\frac{|D^0\rangle-|\dbarzero\rangle}{\sqrt{2}}.
\end{equation}
\end{linenomath*}
The difference in the effective lifetime between $D$ decays to $CP$
eigenstates and flavor eigenstates can be parameterized by $\ycp$.  In
the absence of direct $CPV$, but allowing for small indirect
$CPV$, we have~\cite{Bergmann:2000id}
\begin{linenomath*}
\begin{equation}
\ycp = \frac{1}{2} \left[y{\cos}\phi
         \left( \left|\frac{q}{p}\right| + \left|\frac{p}{q}\right|\right)
                       - x {\sin}\phi
         \left( \left|\frac{q}{p}\right| - \left|\frac{p}{q}\right|\right)
                   \right].
\end{equation}
\end{linenomath*}
In the absence of $CPV$, one has $|{p}/{q}|=1$ and $\phi=0$, leading to
$\ycp=y$.

Although $\dzero-\dbarzero$ mixing from short-distance physics is
suppressed by the CKM matrix~\cite{CKM1,CKM2} and the GIM
mechanism~\cite{GIM}, sizeable charm mixing can arise from
long-distance processes and new physics~\cite{Bianco,Browder}.
Current experimental precision~\cite{HFAG} is not sufficient to
conclude whether physics beyond the SM is involved, and further
constraints are needed.  So far, the most precise determination of the
size of the mixing has been obtained by measuring the time-dependent
decay rate in the $D\rightarrow K^{\pm}\pi^{\mp}$
channel~\cite{LHCb,CDF,Belle}.  However, the resulting information on
the mixing parameters $x$ and $y$ is highly correlated. It is important to
access the mixing parameters $x$ and $y$ directly to provide complementary constraints.

In this analysis, we use a time-integrated method to extract $\ycp$,
as proposed in the references~\cite{Gronau,petrov,cheng,Asner}, which
uses threshold $\dzero \dbarzero$ pair production in $\ee
\rightarrow\gamma^{*}\rightarrow\dzero\dbarzero$.  In this process,
the $\dzero \dbarzero$ pair is in a state of definite $C=-1$, such
that the two $D$ mesons necessarily have opposite $CP$
eigenvalues. 
The method utilizes the semileptonic decays of $D$ meson and
hence, avoids the complications from hadronic effects in $D$ decays,
thus provides a clean and unique way to probe the $\dzero-\dbarzero$
oscillation.

\subsection{Formalism}
In the semileptonic decays of neutral $D$ mesons (denoted as
$D\rightarrow l$)\footnote{Charge-conjugate modes are implied.}, the
partial decay width is only sensitive to flavor content and does not
depend on the $CP$ eigenvalue of the parent $D$ meson.  However, the
total decay width of the $D_{CP\pm}$ does depend on its $CP$
eigenvalue: $\Gamma_{CP\pm} = \Gamma(1\pm y_{CP})$.  Thus, the
semileptonic branching fraction of the $CP$ eigenstates $D_{CP\pm}$ is
$\br{D_{CP\pm}\to l} \approx \br{D\to l} ( 1 \mp y_{CP})$, and $\ycp$
can be obtained as
\begin{linenomath*}
\begin{equation}
\ycp \approx \frac {1}{4} \left(
   \frac{\br{D_{CP-}\to l}}{\br{D_{CP+}\to l}} -
   \frac{\br{D_{CP+}\to l}}{\br{D_{CP-}\to l}} \right). \label{eq:ycp_cal}
\end{equation}
\end{linenomath*}

At BESIII, quantum-correlated $\dzero\dbarzero$ pairs produced at
threshold allow us to measure $\br{D_{CP\pm}\to l}$.  Specifically, we
begin with a fully reconstructed $D$ candidate decaying into a $CP$
eigenstate, the so-called Single Tag (ST).  We have thus tagged the
$CP$ eigenvalue of the partner $D$ meson.  For a subset of the ST
events, the so-called Double Tag (DT) events, this tagged partner $D$
meson is also observed via one of the semileptonic decay channels.
$CP$ violation in $D$ decays is known to be very small~\cite{HFAG},
and can be safely neglected.  Therefore, $\br{D_{CP\mp}\to l}$ can be
obtained as
\begin{linenomath*}
\begin{equation}
\br{D_{CP\mp}\to l} = \frac{N_{CP\pm;l} }{N_{CP\pm}}\cdot\frac{ \varepsilon_{CP\pm}}{ \varepsilon_{CP\pm;l}},
\end{equation}
\end{linenomath*}
where $N_{CP\pm}$ ($N_{CP\pm;l}$) and $\varepsilon_{CP\pm}$
($\varepsilon_{CP\pm;l}$) denote the signal yields and detection
efficiencies of ST decays $D\rightarrow CP\pm$ (DT decays
$D\overline{D}\rightarrow CP\pm ; l$), respectively.  For $CP$
eigenstates, as listed in Table~\ref{table:modes}, we choose modes
with unambiguous $CP$ content and copious yields.  The $CP$ violation
in $K_{S}^{0}$ decays is known to be very small, it is therefore
neglected.  The semileptonic modes used for the DT in this analysis
are $K^\mp e^\pm \nu$ and $K^\mp \mu^\pm \nu$.
\begin{table}
\centering
\caption{$D$ final states reconstructed in this analysis.}
\begin{tabular}{cc}
\toprule
\hline
Type  &   Mode \\
\midrule
$CP+$  & $K^{+}K^{-}$, $\pi^{+}\pi^{-}$, $K^{0}_{S}\pi^{0}\pi^{0}$ \\
$CP-$  & $K_{S}^{0}\pi^{0}$, $K_{S}^{0}\omega$, $K_{S}^{0}\eta$ \\
Semileptonic & $K^\mp e^\pm \nu$, $K^\mp \mu^\pm \nu$  \\
\hline
\bottomrule
\end{tabular}
\label{table:modes}
\end{table}

\subsection{The BESIII detector and data sample}
The analysis presented in this paper is based on a data sample with
an integrated luminosity of 2.92\,fb$^{-1}$~\cite{Lum.} collected with
the BESIII detector~\cite{BESIII} at the center-of-mass energy of
$\sqrt{s}=3.773$\,GeV.  The BESIII detector is a general-purpose
solenoidal detector at the BEPCII~\cite{BEPCII} double storage
rings. The detector has a geometrical acceptance of 93\% of the full solid
angle.  We briefly describe the components of BESIII from the
interaction point (IP) outwards.  A small-cell main drift chamber
(MDC), using a helium-based gas to measure momenta and specific
ionizations of charged particles, is surrounded by a time-of-flight
(TOF) system based on plastic scintillators that determines the
flight times of charged particles.  A CsI(Tl) electromagnetic
calorimeter (EMC) detects electromagnetic showers. These components
are all situated inside a superconducting solenoid magnet, that provides a
1.0\,T magnetic field parallel to the beam direction.  Finally, a
multi-layer resistive plate counter system installed in the iron flux
return yoke of the magnet is used to track muons.  The momentum
resolution for charged tracks in the MDC is 0.5\% for a transverse
momentum of 1\,GeV/$c$. The energy resolution for showers in the EMC is
2.5\% (5.0\%) for 1\,GeV photons in the barrel (end cap) region.  More
details on the features and capabilities of BESIII can be found
elsewhere~\cite{BESIII}.

High-statistics Monte Carlo (MC) simulations are used to evaluate the
detection efficiency and to understand backgrounds.  The {\sc
  geant4}-based~\cite{Geant4} MC simulation program is designed to
simulate interactions of particles in the spectrometer and the
detector response.  For the production of $\psi(3770)$, the {\sc
  kkmc}~\cite{kkmc} package is used; the beam energy spread and the
effects of initial-state radiation (ISR) are included.  The MC samples consist of
the $D\overline{D}$ pairs with consideration of quantum coherence for
all modes relevant to this analysis, non-$D\overline{D}$ decays of
$\psi\rm(3770)$, ISR production of low-mass $\psi$ states, and QED and
$q\bar{q}$ continuum processes.  The effective luminosity of the MC
samples is about 10 times that of the analyzed data.  Known decays
recorded by the Particle Data Group (PDG)~\cite{PDG} are generated
with {\sc evtgen}~\cite{eventgen2,eventgen} using PDG branching
fractions, and the remaining unknown decays are generated with {\sc
  lundcharm}~\cite{lundcharm}.  Final-state radiation (FSR) of charged
tracks is taken into account with the {\sc photos}
package~\cite{photons}.

\section{Event selection and data analysis}\label{sec:Selection}
Each charged track is required to satisfy $|\cos\theta|<0.93$, where
$\theta$ is the polar angle with respect to the beam axis.  Charged
tracks other than $K^0_S$ daughters are required to be within 1\,cm of
the IP transverse to the beam line and within 10\,cm of the IP along
the beam axis.  Particle identification for charged hadrons $h$ ($h =
\pi, K $) is accomplished by combining the measured energy loss
(\dEdx) in the MDC and the flight time obtained from the TOF to form
a likelihood $\mathcal{L}$($h$) for each hadron hypothesis.  The
$K^\pm$ ($\pi^\pm$) candidates are required to satisfy
$\mathcal{L}(K)>\mathcal{L}(\pi)$ ($\mathcal{L}(\pi)>
\mathcal{L}(K)$).

The $K^0_S$ candidates are selected with a vertex-constrained fit from
pairs of oppositely charged tracks, which are required to be within
20\,cm of the IP along the beam direction; no constraint in the
transverse plane is required. The two charged tracks are not subjected
to the particle identification discussed above, and are assumed to be
pions.
We impose $0.487\,\gevcc<M_{\pi^+\pi^{-}}<0.511\,\gevcc$,
that is within about 3 standard deviations of the observed $K^0_S$ mass,
and the two tracks are constrained to originate from a common decay
vertex by requiring the $\chi^2$ of the vertex fit to be less than 100.
The decay vertex is required to be separated from the IP with
a significance greater than two standard deviations.

Reconstructed EMC showers that are separated from the extrapolated
positions of any charged tracks by more than 10 standard deviations
are taken as photon candidates.  The energy deposited in nearby TOF
counters is included to improve the reconstruction efficiency and
energy resolution.  Photon candidates must have a minimum energy of
25\,MeV for barrel showers ($|\cos\theta|<0.80$) and 50\,MeV for end
cap showers ($0.84<|\cos\theta|<0.92$). The showers in the gap between
the barrel and the end cap regions are poorly reconstructed and thus
excluded.  The shower timing is required to be no later than 700\,ns
after the reconstructed event start time to suppress electronic noise
and energy deposits unrelated to the event.  The $\eta$ and $\pi^0$
candidates are reconstructed from pairs of photons.  Due to the poorer
resolution in the EMC end cap regions, those candidates with both
photons coming from EMC end caps are rejected.  The invariant mass
$M_{\gamma\gamma}$ is required to be
$0.115\,\gevcc<M_{\gamma\gamma}<0.150\,\gevcc$ for $\pi^0$ and
$0.505\,\gevcc<M_{\gamma\gamma}<0.570\,\gevcc$ for $\eta$ candidates.
The photon pair is kinematically constrained to the nominal mass of
the $\pi^0$ or $\eta$~\cite{PDG} to improve the meson four-vector
calculation.

\begin{figure}
\begin{center}
\includegraphics[width=0.45\textwidth]{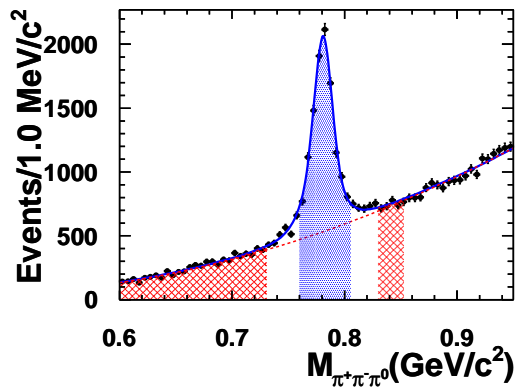}
\caption{Fit to the invariant mass $\unit{M}_{\pi^+\pi^-\pi^0}$ for events reconstructed from data.
The solid line is the total fit and the dashed line shows the polynomial background.
The shaded area shows the signal region and cross-hatched areas show the sidebands.}
\label{fig:omega}
\end{center}
\end{figure}

The $\omega$ candidates are reconstructed through the decay $\omega
\to \pi^{+}\pi^{-}\pi^{0}$.  For all modes with $\omega$ candidates,
sideband events in the $M_{\pi^+\pi^-\pi^0}$ spectrum are used to
estimate peaking backgrounds from non-$\omega$ $D\rightarrow
K_{S}^0\pi^{+}\pi^{-}\pi^{0}$ decays.  We take the signal region as
(0.7600, 0.8050)\,$\gevcc$ and the sideband regions as (0.6000,
0.7300)\,$\gevcc$ or (0.8300, 0.8525)\,$\gevcc$.  The upper edge of
the right sideband is restricted because of the $K^*\rho$ background
that alters the shape of $M_{\pi^+\pi^-\pi^0}$. The sidebands are
scaled to the estimated peaking backgrounds in the signal region. The
scaling factor is determined from a fit to the
$M_{\pi^{+}\pi^{-}\pi^{0}}$ distribution in data, as shown in
Fig.~\ref{fig:omega}, where the $\omega$ signal is determined with the
MC shape convoluted with a Gaussian whose parameters are left free in
the fit to better match data resolution, and the background is modeled
by a polynomial function.
\subsection{Single tags using $CP$ modes}
To identify the reconstructed $D$ candidates, we use two variables,
the beam-constrained mass $\mbc$ and the energy difference $\delE$,
which are defined as
\begin{linenomath*}
\begin{equation}
\mbc \equiv \sqrt{E_{\rm beam}^{2}/c^{4}-|\vec{p}_{D}|^{2}/c^{2}},
\end{equation}
\begin{equation}
\Delta E\equiv E_{D}-E_{\rm beam},
\end{equation}
\end{linenomath*}
where $\vec{p}_{{D}}$ and ${E}_{D}$ are the momentum and energy of the
$D$ candidate in the $e^+e^-$ center-of-mass system, and $E_{\rm
  beam}$ is the beam energy.  The $D$ signal peaks at the nominal $D$
mass in $\mbc$ and at zero in $\delE$.  We accept only one candidate
per mode per event; when multiple candidates are present, the one with
the smallest $|\delE|$ is chosen. Since the correlation between $\delE$ and $\mbc$ is found to be small,
this will not bias the background distribution in $\mbc$.
We apply the mode-dependent $\delE$
requirements listed in Table~\ref{tab:deltaE_cuts}.

For $K^{+}K^{-}$ and $\pi^{+}\pi^{-}$ ST modes, if candidate events
contain only two charged tracks, the following requirements are
applied to suppress backgrounds from cosmic rays and Bhabha events.
First, we require at least one EMC shower separated from the tracks of
the ST with energy larger than 50\,MeV.  Second, the two ST tracks
must not be both identified as muons or electrons, and, if they have
valid TOF times, the time difference must be less than 5\,ns.  Based
on MC studies, no peaking background is present in $\mbc$ in our ST
modes except for the $K_{S}^{0}\pi^0$ mode.  In the $K_{S}^{0}\pi^0$
ST mode, there are few background events from $D^0\to \rho\pi$. From
MC studies, the estimated fraction is less than $0.5\%$; this will be
considered in the systematic uncertainties.

\begin{table}
\caption{Requirements on $\Delta E$ for ST $D$ \rm candidates.}
\centering
\begin{tabular}{cc}
\toprule
\hline
Mode  &   Requirement (GeV) \\
\midrule
$K^{+}K^{-}$               & $-0.020 < \Delta E < 0.020$ \\
$\pi^{+}\pi^{-}$           & $-0.030 < \Delta E < 0.030$ \\
$K_{S}^{0}\pi^{0}\pi^{0} $ & $-0.080 < \Delta E < 0.045$ \\
$K_{S}^{0}\pi^{0}$         & $-0.070 < \Delta E < 0.040$ \\
$K_{S}^{0}\omega$          & $-0.050 < \Delta E < 0.030$ \\
$K_{S}^{0}\eta$            & $-0.040 < \Delta E < 0.040$ \\
\hline
\bottomrule
\end{tabular}
\label{tab:deltaE_cuts}
\end{table}

\begin{figure*}
\begin{center}
\includegraphics[width=0.9\linewidth]{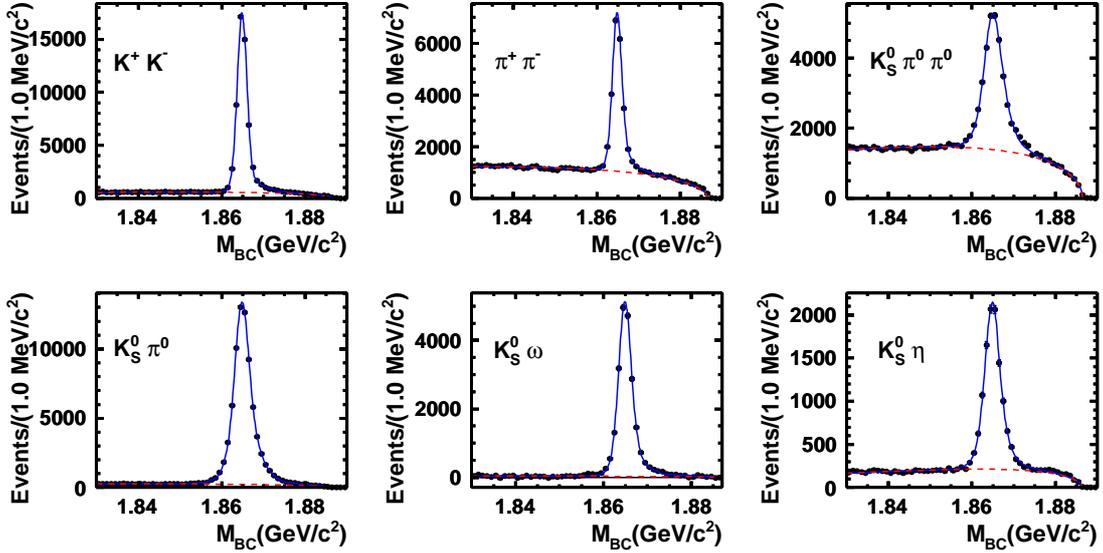}
\caption{The $\mbc$ distributions for ST $D$ candidates from data.
The solid line is the total fit and the dashed line shows the background contribution described by an ARGUS function.}
\label{fig:mbcfits}
\end{center}
\end{figure*}

The $\mbc$ distributions for the six ST modes are shown in
Fig.~\ref{fig:mbcfits}.  Unbinned maximum likelihood fits are
performed to obtain the numbers of ST yields except in the
$K^{0}_{S}\omega$ mode, for which a binned least-square fit is applied
to the $\mbc$ distribution after subtraction of the $\omega$
sidebands.  In each fit, the signal shape is derived from simulated
signal events convoluted with a bifurcated Gaussian with free
parameters to account for imperfect modeling of the detector
resolution and beam energy calibration. Backgrounds are described by
the ARGUS~\cite{Argus} function.  The measured ST yields in the signal
region of 1.855\,GeV/$c^2<\mbc<$1.875\,GeV/$c^2$ and the corresponding
efficiencies are given in Table~\ref{table:yields}.

\begin{table}
\centering
\caption{
Yields and efficiencies of all ST and DT modes, where $N_{CP\pm}$ ($N_{CP\pm;l}$) and $\varepsilon_{CP\pm}$ ($\varepsilon_{CP\pm;l}$) denote signal yields and detection efficiencies of $D\rightarrow CP\pm$ ($D\overline{D}\rightarrow CP\pm ; l$), respectively. The uncertainties are statistical only.}
\begin{tabular}{ccc}
\toprule
\hline
 ST Mode  &  $N_{CP\pm}$  & $\varepsilon_{CP\pm}$ (\%) \\
\midrule
$K^{+}K^{-}$                    &  54494 $\pm$ 251  & 61.32 $\pm$ 0.18 \\
$\pi^{+}\pi^{-}$                &  19921 $\pm$ 174  & 64.09 $\pm$ 0.18 \\
$K_{S}^{0}\pi^{0}\pi^{0}$       &  24015 $\pm$ 236  & 16.13 $\pm$ 0.08 \\
$K_{S}^{0}\pi^{0}$              &  71421 $\pm$ 285  & 40.67 $\pm$ 0.14 \\
$K_{S}^{0}\omega$               &  20989 $\pm$ 243  & 13.44 $\pm$ 0.07 \\
$K_{S}^{0}\eta$                 &  9878  $\pm$ 117  & 34.39 $\pm$ 0.13 \\
\midrule
DT Mode      &   $N_{CP\pm;l}$ &  $\varepsilon_{CP\pm;l}$ (\%) \\
\midrule
$K^{+}K^{-}$, $K e \nu$                &  1216 $\pm$ 40 & 39.80 $\pm$ 0.14 \\
$\pi^{+}\pi^{-}$, $K e \nu$            &  427  $\pm$ 23 & 41.75 $\pm$ 0.14 \\
$K_{S}^{0}\pi^{0}\pi^{0}$, $K e \nu$   &  560  $\pm$ 28 & 11.05 $\pm$ 0.07 \\
$K_{S}^{0}\pi^{0}$, $K e \nu$          &  1699 $\pm$ 47 & 26.70 $\pm$ 0.12 \\
$K_{S}^{0}\omega$,  $K e \nu$          &  481  $\pm$ 30 & 9.27  $\pm$ 0.07 \\
$K_{S}^{0}\eta$, $K e \nu$             &  243  $\pm$ 17 & 22.96 $\pm$ 0.11 \\
$K^{+}K^{-}$, $K \mu \nu$              &  1093 $\pm$ 37 & 36.89 $\pm$ 0.14 \\
$\pi^{+}\pi^{-}$, $K \mu \nu$          &  400  $\pm$ 23 & 38.43 $\pm$ 0.15 \\
$K_{S}^{0}\pi^{0}\pi^{0}$, $K \mu \nu$ &  558  $\pm$ 28 & 10.76 $\pm$ 0.08 \\
$K_{S}^{0}\pi^{0}$, $K \mu \nu$        &  1475 $\pm$ 43 & 25.21 $\pm$ 0.12 \\
$K_{S}^{0}\omega$, $K \mu \nu$         &  521  $\pm$ 27 & 8.75  $\pm$ 0.07 \\
$K_{S}^{0}\eta$, $K \mu \nu$           &  241  $\pm$ 18 & 21.85 $\pm$ 0.11 \\
\hline
\bottomrule
\end{tabular}
\label{table:yields}
\end{table}

\subsection{Double tags of semileptonic modes}
In each ST event, we search among the unused tracks and showers for
semileptonic $D\rightarrow K e (\mu) \nu$ candidates.  We require that
there be exactly two oppositely-charged tracks that satisfy the
fiducial requirements described above.

In searching for $K\mu\nu$ decays, kaon candidates are required to
satisfy $\mathcal{L}(K)>\mathcal{L}(\pi)$. If the two tracks can pass
the criterion, the track with larger $\mathcal{L}(K)$ is taken as the
$K^{\pm}$ candidate, and the other track is assumed to be the $\mu$
candidate.  The energy deposit in the EMC of the $\mu$ candidate is
required to be less than 0.3\,GeV.  We further require the $K\mu$
invariant mass $M_{K\mu}$ to be less than 1.65\,$\gevcc$ to reject
$D\rightarrow K\pi$ backgrounds. The total energy of remaining
unmatched EMC showers, denoted as $E_{\rm extra}$, is required to be
less than 0.2\,GeV to suppress $D\rightarrow K\pi\pi^0$ backgrounds.
To reduce backgrounds from the $D\rightarrow K e \nu$ process, the
ratio $\mathcal{R}_{\mathcal{L}^{'}}(e)\equiv
\mathcal{L}^{'}(e)/[\mathcal{L}^{'}(e) + \mathcal{L}^{'}(\mu)
+\mathcal{L}^{'}(\pi) + \mathcal{L}^{'}(K)]$ is required to be less
than 0.8, where the likelihood $\mathcal{L}^{'}(i)$ for the hypothesis
$i= e$, $\mu$, $\pi$ or $K$ is formed by combining EMC information
with the \dEdx and TOF information.

To select $Ke\nu$ events, electron candidates are required to satisfy
$\mathcal{L}^{'}(e)>0.001$ and
$\mathcal{R}^{'}_{\mathcal{L}^{'}}(e)>$0.8, where
$\mathcal{R}^{'}_{\mathcal{L}^{'}}(e)\equiv
\mathcal{L}^{'}(e)$/[$\mathcal{L}^{'}(e)+\mathcal{L}^{'}(\pi)+\mathcal{L}^{'}(K)$].
If both tracks satisfy these requirements, the one with larger
$\mathcal{R}^{'}_{\mathcal{L}^{'}}(e)$ is taken as the electron.  The
remaining track is required to satisfy
$\mathcal{L}(K)>\mathcal{L}(\pi)$.

The variable $U_{\rm miss}$ is used to distinguish semileptonic signal events from background:
\begin{linenomath*}
\begin{equation}
U_{\rm miss}\equiv E_{\rm miss}-c|\vec{p}_{\rm miss}|,
\end{equation}
\end{linenomath*}
where,
\begin{linenomath*}
\begin{equation}
E_{\rm miss} \equiv E_{\rm beam}-E_{K}-E_{l},
\end{equation}
\begin{equation}
 \vec{p}_{\rm miss} \equiv - \left[
     \vec{p}_K + \vec{p}_l
               + \hat{p}_{\rm ST}\sqrt{E_{\rm beam}^2/c^{2}-c^{2}m_D^2}
                             \right],
\end{equation}
\end{linenomath*}
$E_{K(l)}$ ($\vec{p}_{K(l)}$) is the energy (three-momentum) of
$K^\mp$ ($l^\pm$), $\hat{p}_{\rm ST}$ is the unit vector in the
direction of the reconstructed $CP$-tagged $D$ and $m_D$ is the
nominal $D$ mass.  The use of the beam energy and the nominal $D$ mass
for the magnitude of the $CP$-tagged $D$ improves the $\umiss$
resolution.  Since $E$ equals to $|\vec{p}|c$ for a neutrino, the signal peaks at
zero in $U_{\rm miss}$.

\begin{figure*}
\begin{center}
\includegraphics[width=0.9\linewidth]{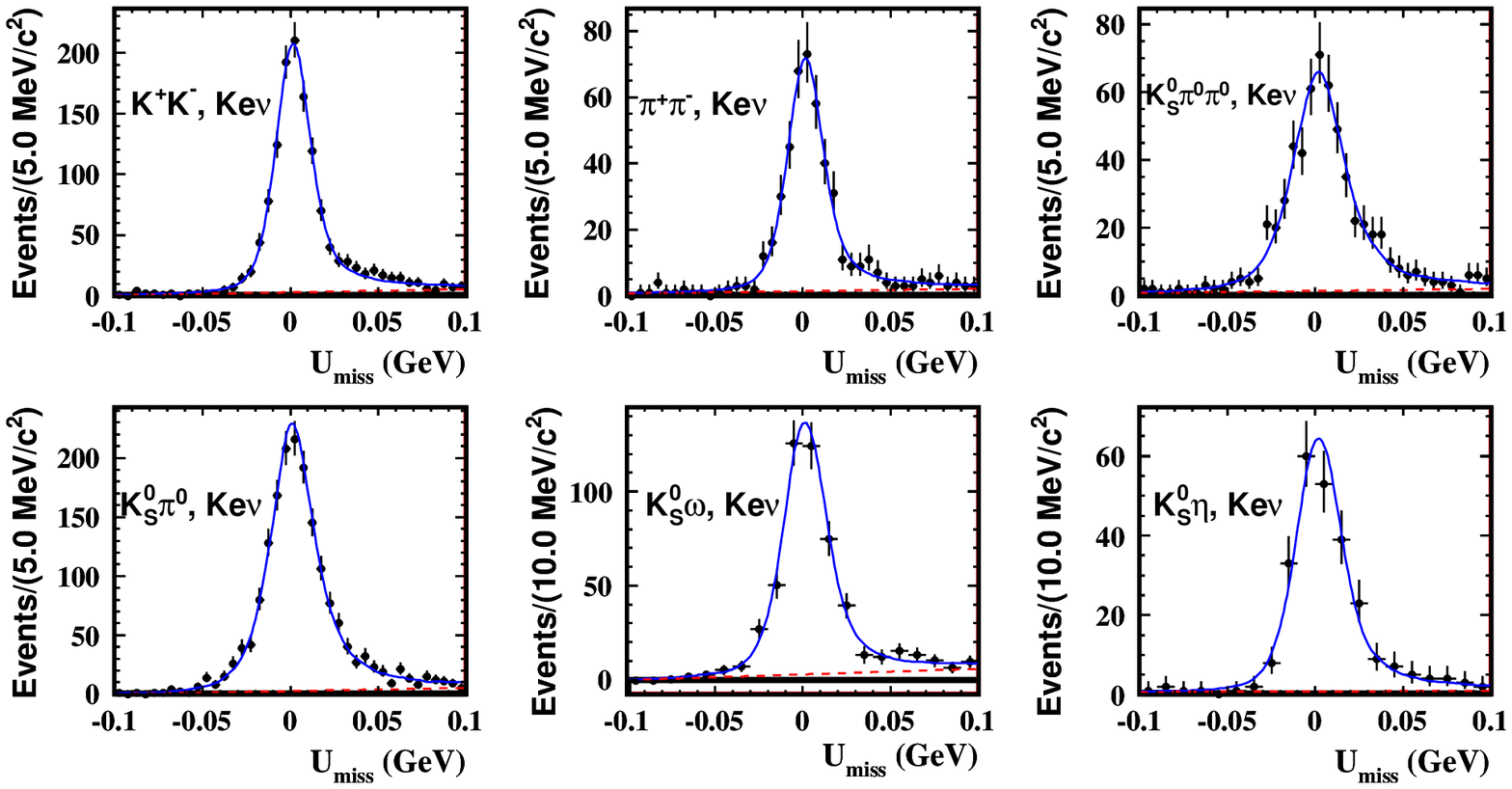}
\includegraphics[width=0.9\linewidth]{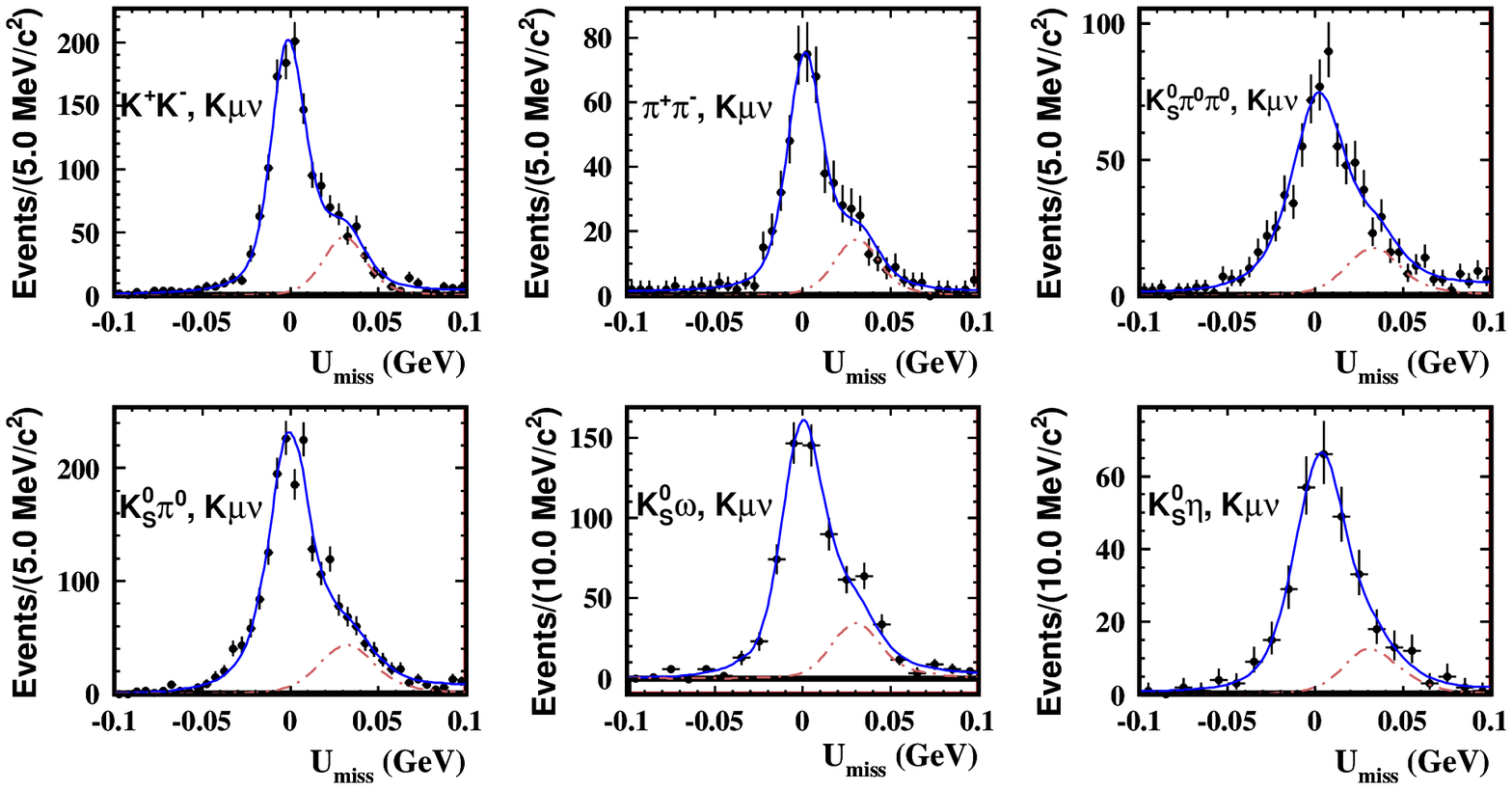}
\caption{Fit to the $U_{\rm miss}$ distributions for selected DT events from data.
In each plot, the solid line is the total fit, the dashed line in $Ke\nu$
shows the contribution of polynomial backgrounds, and the dash-dotted line
in $K\mu\nu$ shows the contribution of the main $K\pi\pi^0$ backgrounds.}
\label{fig:ufit}
\end{center}
\end{figure*}

The $U_{\rm miss}$ distributions are shown in Fig.~\ref{fig:ufit},
where the tagged-$D$ is required to be in the region of
1.855\,GeV/$c^2<\mbc<$1.875\,GeV/$c^2$.  DT yields, obtained by
fitting the $\umiss$ spectra, are listed in Table~\ref{table:yields}.
Unbinned maximum likelihood fits are performed for all modes except
for modes including $\omega$.  For modes including an $\omega$, binned
least-square fits are performed to the $\omega$ sideband-subtracted
$\umiss$ distributions.  In each fit, the $K e \nu$ or $K\mu\nu$
signal is modeled by the MC-determined shape convoluted with a
bifurcated Gaussian where all parameters are allowed to vary in the
fit.  Backgrounds for $K e \nu$ are well described with a first-order
polynomial.  However, in the $K \mu \nu$ mode, backgrounds are more
complex and consist of three parts.  The primary background comes from
$D\rightarrow K\pi\pi^0$ decay.  To better control this background, we
select a sample of $D\rightarrow K\pi\pi^0$ in data by requiring
$E_{\rm extra}>$0.5\,GeV, in which the $U_{\rm miss}$ shape of
$K\pi\pi^0$ is proved to be basically the same as that in the region
of $E_{\rm extra}<$0.2\,GeV in MC simulation.  The selected
$K\pi\pi^{0}$ sample is used to extract the resolution differences in
the $U_{\rm miss}$ shape of $K\pi\pi^0$ in MC and data, and to obtain
the $D\rightarrow K\pi\pi^0$ yields in $E_{\rm extra}>$0.5\,GeV
region.  Then, in fits to $U_{\rm miss}$, the $K\pi\pi^{0}$ is
described by the resolution-corrected shape from MC simulations and
its size is fixed according to the relative simulated efficiencies of
the $E_{\rm extra}>$0.5\,GeV and $E_{\rm extra}<$0.2\,GeV selection
criteria.  The second background from $Ke\nu$ events is modeled by a
MC-determined shape. Its ratio to the signal yields is about 3.5\%
based on MC studies and is fixed in the fits.  Background in the third
category includes all other background processes, which are well
described with a first-order polynomial.

\section{Systematic uncertainties}

Most sources of uncertainties for the ST or DT efficiencies, such as
tracking, PID, and $\pi^{0}$, $\eta$, $K_{S}^{0}$ reconstruction,
cancel out in determining $\ycp$.  The main systematic uncertainties
come from the background veto, modeling of the signals and
backgrounds, fake tagged signals, and the $CP$-purity of ST events.

The cosmic and Bhabha veto is applied only for the $KK$ and $\pi\pi$
ST events which have only two tracks.  The effect of this veto is
estimated based on MC simulation.  We compare the cases with and
without this requirement and the resultant relative changes in ST
efficiencies are about 0.3\% for both the $KK$ and $\pi\pi$ modes.
The resulting systematic uncertainty on $\ycp$ is 0.001.

Peaking backgrounds are studied for different ST modes, especially for
$\rho\pi$ backgrounds in the $K^0_{S}\pi^{0}$ tag mode and
$K^0_{S}\pi^{+}\pi^{-}\pi^{0}$ backgrounds in the $K^0_{S}\omega$ tag
mode.  Based on a study of the inclusive MC samples, the fraction of
peaking backgrounds in $K^0_{S}\pi^{0}$ is 0.3\%. The uncertainties on
$\ycp$ caused by this is about 0.001.  Uncertainties from the sideband
subtraction of peaking backgrounds for the $K^0_{S} \omega$ mode are
studied by changing the sideband and signal regions; changes in the
efficiency-corrected yields are negligible.

Fits to the $\mbc$ and $\umiss$ spectra could induce systematic errors
by the modeling of the signal and background shape.  The MC-determined
signal shapes convoluted with a Gaussian are found to describe the
data well, and systematic uncertainties from the modeling of the
signal are assumed to be negligible.  To estimate uncertainties from
modeling of backgrounds, different methods are considered. For the
$CP$ ST yields, we include an additional background component to
account for the $\psi({\rm{3770}})\rightarrow D\overline{D}$ process
with a shape determined by MC simulation whose yield is determined in
the fit.  The uncertainties in the fits to $\mbc$ are uncorrelated
among different tag modes, and the obtained change on $\ycp$ is 0.001.
For the DT semileptonic yields, the polynomial functions that are
used to describe backgrounds in our nominal fits are replaced by a
shape derived from MC simulations.  For the $K\mu\nu$ mode, the size
of the main background $K\pi\pi^{0}$ is fixed in our nominal fit, so
the statistical uncertainties of the number of selected $K\pi\pi^{0}$
events introduces a systematic error.  To estimate the associated
uncertainty, we vary its size by $\pm$1 standard deviation based on
the selected $K\pi\pi^{0}$ samples.  Systematic uncertainties due to
the $\umiss$ fits are treated as positively correlated among different
tag modes.  We take the maximum change on the resultant $\ycp$, that
is 0.006, as systematic uncertainty.

The DT yields are obtained from the fit to the $\umiss$ spectra.
However, one also has to consider events that peak at $\umiss$ but
are backgrounds in the $\mbc$ spectra, the so-called fake tagged
signals.  This issue is examined by fitting to the $\mbc$ versus
$\umiss$ two-dimensional plots.  From this study, the fake tagged
signal component is proved to be very small.  The resulting difference
on $\ycp$ is 0.002 and assigned as a systematic uncertainty.

We study the $CP$-purities of ST modes by searching for same-$CP$ DT
signals in data.  Assuming $CP$ conservation in the charm sector, the
same-$CP$ process is prohibited, unless the studied $CP$ modes are not
pure or the initial $C$-odd $\dzero \dbarzero$ system is diluted.  The
$CP$ modes involving $K_{S}^{0}$ are not pure due to the existence of
small $CPV$ in $K^{0}-\overline{K}{}^{0}$ mixing~\cite{PDG}.  However,
this small effect is negligible with our current sensitivity.  Hence,
$K_{S}^{0}\pi^{0}$ is assumed to be a clean $CP$ mode, as its
background level is very low.  As a conservative treatment, we study
DT yields of ($K_{S}^{0}\pi^0$, $K_{S}^{0}\pi^0$) to verify its pure
$CP$-odd eigenstate nature and the $CP$-odd environment of the $\dzero
\dbarzero$ pair.  The observed numbers of this DT signals are quite
small, and we estimate the dilution of the $C$-odd initial state to be
less than 2\% at 90\% confidence level. This affects our measurement
of $\ycp$ by less than 0.001.  The purity of the $K_{S}^{0}\pi^0$ mode
is found to be larger than 99\%.  Due to the complexity of the
involved non-resonant and resonant processes in
$K_{S}^{0}\pi^{0}\pi^{0}$ and $K_{S}^{0}\omega$, the $CP$-purities of
these tag modes could be contaminated.  We take the mode $K^+K^-$ as a
clean $CP$-even tag to test $K_{S}^{0}\pi^{0}\pi^{0}$, and take
$K_{S}^{0}\pi^0$ to test $K_{S}^{0}\omega$ and $K_{S}^{0}\eta$.  The
$CP$-purities of $K_{S}^{0}\pi^0\pi^{0}$, $K_{S}^{0}\omega$ and
$K_{S}^{0}\eta$ are estimated to be larger than 89.4\%, 93.3\% and
93.9\%, respectively.  Based on the obtained $CP$ purities, the
corresponding maximum effect on the determined $\ycp$ is assigned as
systematic uncertainty.

Systematic uncertainties from different sources are assumed to be independent and are combined in quadrature to obtain the overall $\ycp$ systematic uncertainties. The resultant total $\ycp$ systematic uncertainties is 0.007.

\begin{table*}
\centering
\caption{Summary of systematic uncertainties. Relative systematic uncertainties are listed for each tag mode in percent, while the resulting absolute uncertainties on $y_{cp}$ are shown in the last column.
Negligible uncertainties are denoted by ``--''.}
~\label{tab:sys_summary}
\begin{tabular}{ccccccc|c}
\toprule
\hline
             & $K^{+}K^{-}$  &  $\pi^{+}\pi^{-}$ & $K_{S}^{0}\pi^{0}\pi^{0}$   & $K_{S}^{0}\pi^{0}$ &  $K_{S}^{0}\omega$   & $K_{S}^{0}\eta$  & $y_{cp}$          \\
\hline
Background              & 0.3  &       0.3             &  -                     & 0.3                &    --                &  --        & 0.001                  \\
\hline
$M_{BC}$ fit            & 0.4  &       0.1             & 2.4                    & 0.4                &   0.1                &  1.4       & 0.001                  \\
\hline
$\umiss$ fit ($Ke\nu$)  & 1.8  &       1.3             & 2.4                    & 1.6                &   8.1                &  1.2       & \multirow{2}{*}{0.006} \\
$\umiss$ fit ($K\mu\nu$)& 3.2  &       7.0             & 4.6                    & 2.5                &   1.7                &  1.7       &                        \\
\hline
Fake tag ($Ke\nu$)      & 0.2  &       1.4             & 0.9                    & 1.2                &   3.1                &  0.4       & \multirow{2}{*}{0.002} \\
Fake tag ($K\mu\nu$)    & 1.0  &       0.7             & 0.5                    & 0.9                &   4.8                &  0.4       &                        \\
\hline
$CP$-purity             & --   &        --             & 0.4                    & --                 &   0.2                &  0.2       &  0.001                 \\
\hline
\bottomrule
\end{tabular}
\end{table*}

\section{Results}
The branching ratios of $K^\mp e^\pm \nu$ and $K^\mp \mu^\pm \nu$ are
summed to obtain $\br{D_{CP\mp}\to l} = \br{D_{CP\mp}\to K e \nu} +
\br{D_{CP\mp}\to K \mu \nu}$.  To combine results from different $CP$
modes, the standard weighted least-square method is
utilized~\cite{PDG}.  The weighted semileptonic branching fraction
$\tilde{\mathcal{B}}_{D_{CP\pm}\rightarrow l}$ is determined by
minimizing
\begin{linenomath*}
\begin{equation}
\chi^{2}= \sum_\alpha \frac{ \left(
       \tilde{\mathcal{B}}_{D_{CP\pm}\rightarrow l} -
             {\mathcal{B}}^{\alpha}_{D_{CP\pm}\to l} \right)^{2}}
        { \left(\sigma^{\alpha}_{CP\pm} \right)^{2}},
\end{equation}
\end{linenomath*}
where $\alpha$ denotes different $CP$-tag modes and
$\sigma^{\alpha}_{CP\pm}$ is the statistical error of
${\mathcal{B}^{\alpha}_{D_{CP\pm}\to l}}$ for the given tag mode.  The
branching fractions of $\br{D_{CP\pm}\to l}$ and the obtained
$\tilde{\mathcal{B}}_{D_{CP\pm}\rightarrow l}$ are listed in
Table~\ref{tab:br}.  Finally, $y_{CP}$ is calculated using
Eq.~\eqref{eq:ycp_cal}, with $\br{D_{CP\pm}\to l}$ replaced by
$\tilde{\mathcal{B}}_{D_{CP\pm}\rightarrow l}$.  We obtain $\ycp
=(-2.0\pm 1.3 \pm 0.7 )\%$, where the first uncertainty is statistical
and the second is systematic.

\begin{table}
\centering
\caption{Values of branching ratio of $D_{CP\pm \to l}$ obtained from different tag modes and the combined branching ratio. The errors shown are statistical only.}
\label{tab:br}
\begin{tabular}{ccccc}
\toprule
\hline
Tag Mode &  ${\mathcal{B}}_{D_{CP-}\rightarrow Ke\nu}$ (\%)  &    ${\mathcal{B}}_{D_{CP-}\rightarrow K\mu\nu}$ (\%) & ${\mathcal{B}}_{D_{CP-}\rightarrow l}$ (\%)   \\
\midrule
$K^{+}K^{-}$       &   3.44  $\pm$ 0.12        &   3.33  $\pm$ 0.12   &  6.77 $\pm$ 0.17 \\
$\pi^{+}\pi^{-}$   &   3.29  $\pm$ 0.18        &   3.35  $\pm$ 0.20   &  6.64 $\pm$ 0.27 \\
$K_{S}^{0}\pi^{0}\pi^{0}$ & 3.40  $\pm$   0.18 &   3.48  $\pm$  0.18  &  6.89 $\pm$ 0.26\\
\midrule
\midrule
$\tilde{\mathcal{B}}_{D_{CP-}\rightarrow l}$   &  3.40 $\pm$ 0.09    &  3.37   $\pm$ 0.09 & 6.77 $\pm$ 0.12 \\
\midrule
Tag Mode &  ${\mathcal{B}}_{D_{CP+}\rightarrow Ke\nu}$ (\%)   &  ${\mathcal{B}}_{D_{CP+}\rightarrow K\mu\nu}$ (\%) & ${\mathcal{B}}_{D_{CP+}\rightarrow l}$ (\%)  \\
\midrule
$K_{S}^{0}\pi^{0}$ &  3.62 $\pm$   0.10   &   3.33 $\pm$  0.10 & 6.96 $\pm$ 0.15  \\
$K_{S}^{0}\omega$  &  3.32 $\pm$   0.21   &   3.81 $\pm$  0.21 & 7.14 $\pm$ 0.30 \\
$K_{S}^{0}\eta$    &  3.68  $\pm$  0.26   &   3.84 $\pm$ 0.29  & 7.52 $\pm$ 0.40\\
\hline
\midrule
$\tilde{\mathcal{B}}_{D_{CP+}\rightarrow l}$  &   3.58 $\pm$   0.09  &   3.46  $\pm$ 0.09  & 7.04 $\pm$ 0.13 \\
\bottomrule
\end{tabular}
\end{table}

\section{Summary}
Using quantum-correlated $\dzero\dbarzero$ pairs produced
at $\sqrt{s}$ = 3.773\,GeV, we employ a
$CP$-tagging technique to obtain the $\ycp$ parameter of $\dzero-\dbarzero$ oscillations.
Under the assumption of no direct $CPV$ in the $D$ sector, we obtain
$\ycp =(-2.0\pm 1.3(\rm stat.)\pm 0.7(\rm syst.))\%$.
This result is compatible with the previous measurements~\cite{HFAG,BaBar,Belle_ycp,LHCb_AGam} within about two standard deviations.
However, the precision is still statistically limited and less precise than the current world average~\cite{PDG}.
Future efforts using a global fit~\cite{LSfit} may better exploit
the BESIII data, leading to a more precise result.

\section{Acknowledgments}
The BESIII collaboration thanks the staff of BEPCII and the IHEP
computing center for their strong support. This work is supported
in part by National Key Basic Research Program of China under
Contract No. 2015CB856700; Joint Funds of the National Natural
Science Foundation of China under Contracts Nos.  11079008, 11179007, U1232201,
U1332201; National Natural Science Foundation of China (NSFC) under
Contracts Nos. 11125525, 11235011, 11275266, 11322544, 11335008, 11425524; the
Chinese Academy of Sciences (CAS) Large-Scale Scientific Facility Program;
CAS under Contracts Nos. KJCX2-YW-N29, KJCX2-YW-N45; 100 Talents Program of CAS;
INPAC and Shanghai Key Laboratory for Particle Physics and Cosmology;
German Research Foundation DFG under Contract No. Collaborative Research Center CRC-1044;
Istituto Nazionale di Fisica Nucleare, Italy; Ministry of Development of Turkey
under Contract No. DPT2006K-120470; Russian Foundation for Basic Research under Contract No. 14-07-91152;
U. S. Department of Energy under Contracts Nos. DE-FG02-04ER41291, DE-FG02-05ER41374,
DE-FG02-94ER40823, DESC0010118; U.S. National Science Foundation; University
of Groningen (RuG) and the Helmholtzzentrum fuer Schwerionenforschung GmbH (GSI),
Darmstadt; WCU Program of National Research Foundation of Korea under Contract
No. R32-2008-000-10155-0.

\bibliographystyle{model1a-num-names}
\bibliography{mybibl}


\end{document}

%% file: authors_jan2015.tex
\author{
  \begin{small}
    \begin{center}
      M.~Ablikim$^{1}$, M.~N.~Achasov$^{8,a}$, X.~C.~Ai$^{1}$,
      O.~Albayrak$^{4}$, M.~Albrecht$^{3}$, D.~J.~Ambrose$^{43}$,
      A.~Amoroso$^{47A,47C}$, F.~F.~An$^{1}$, Q.~An$^{44}$,
      J.~Z.~Bai$^{1}$
% ...
      \\
      \vspace{0.2cm}
      (BESIII Collaboration)\\
      \vspace{0.2cm} {\it
        $^{1}$ Institute of High Energy Physics, Beijing 100049, People's Republic of China\\
% ...
$^{3}$ Bochum Ruhr-University, D-44780 Bochum, Germany\\
        $^{4}$ Carnegie Mellon University, Pittsburgh, Pennsylvania 15213, USA\\
% ...
$^{8}$ G.I. Budker Institute of Nuclear Physics SB RAS (BINP), Novosibirsk 630090, Russia\\
% ...
$^{43}$ University of Rochester, Rochester, New York 14627, USA\\
        $^{44}$ University of Science and Technology of China, Hefei 230026, People's Republic of China\\
% ...
$^{47}$ (A)University of Turin, I-10125, Turin, Italy; (B)University of Eastern Piedmont, I-15121, Alessandria, Italy; (C)INFN, I-10125, Turin, Italy\\
% ...
$^{a}$ Also at the Novosibirsk State University, Novosibirsk, 630090, Russia\\
% ...
      }\end{center}
    \vspace{0.4cm}
  \end{small}
}